\documentclass[12pt]{iopart}

\usepackage{iopams}
\usepackage{color}
\usepackage{cite}
\usepackage[dvips]{graphicx}
\usepackage{bm}

\newcommand{\nm}{\nonumber\\}

\newcommand{\kB}{k_\mathrm{B}}

\newcommand{\ep}{\varepsilon}
\newcommand{\eq}{\mathrm{eq}}

\newcommand{\TL}{T_\mathrm{L}}
\newcommand{\TR}{T_\mathrm{R}}

\newcommand{\subLR}{\mathrm{L/R}}

\begin{document}

\title[Multiplicative Langevin Equation to Reproduce Long-time Properties]{Multiplicative Langevin Equation to Reproduce Long-time Properties of Nonequilibrium Brownian Motion}
\author{Atsumasa Seya$^1$, Tatsuya Aoyagi$^1$, Masato Itami$^{2, a}$, Yohei Nakayama$^{3, b}$ and Naoko Nakagawa$^{1, c}$}
\address{$^1$Department of Physics, Ibaraki University, Mito 310-8512, Japan}
\address{$^2$Fukui Institute for Fundamental Chemistry, Kyoto University, Kyoto 606-8103, Japan}
\address{$^3$Department of Applied Physics, Tohoku University, Sendai 980-8579, Japan}
\eads{\mailto{$^a$itami@fukui.kyoto-u.ac.jp}, 
\mailto{$^b$r\_nakayama@tohoku.ac.jp},
\mailto{$^c$naoko.nakagawa.phys@vc.ibaraki.ac.jp}}
\date{\today}

\begin{abstract}
We statistically examine long time sequences of Brownian motion for a nonequilibrium version of the Rayleigh piston model and confirm
that the third cumulant of a long-time displacement for the nonequilibrium Brownian motion linearly increases with the observation time interval.
We identify a multiplicative Langevin equation that can reproduce the cumulants  
of the long-time displacement up to at least the third order, as well as its mean, variance and skewness. 
The identified Langevin equation involves a velocity-dependent friction coefficient that breaks the time-reversibility and may act as a generator of the directionality.
Our method to find the Langevin equation is not specific to the Rayleigh piston model but may be applied to a general time sequence in various fields. 
\vspace*{1ex}
\end{abstract}

\section{Introduction}

Brownian motion refers to random fluctuating phenomena ubiquitously observed in nature.
Typically, it is observed as trajectories of molecules, colloidal particles, biomotors in cells, agents in active matter, stars within galaxies, interfaces, market prices and so on \cite{Svoboda,Noji,Paxton,Jiang,stars,Takeuchi,Ide,Bouchaud}.
These dynamics are often 
described by the Langevin equation.
Once we obtain the equation, we can predict unknown properties and propose experimental methods to estimate previously unmeasured quantities from the theoretical point of view.
Indeed, the stochastic energetics \cite{Sekimoto-book,Seifert-review} based on the Langevin equation was developed to analyze the Brownian motion of molecular machines and enables us to approach their design principles.
For example, a nonequilibrium equality \cite{Harada-Sasa} revealed that a rotary motor protein dissipates free energy at almost 100\% through its rotational motion \cite{ToyabeETAL}, but a processive motor protein wastes in parts other than the translational motion \cite{Ariga}.
Thus, the Langevin equation forms the basis of theoretical investigations that are inseparable from developments of experimental techniques.

The identification of the Langevin equation in equilibrium is well-established because its form is strictly restricted by the fluctuation-dissipation theorem \cite{Risken-book,vanKampen-book}.
Conventionally, nonequilibrium Brownian motion has been described by adding terms representing nonequilibrium effects to the Langevin equations for the equilibrium Brownian motion.
However, these expressions with the fluctuation-dissipation theorem are not justified out of equilibrium \cite{Sekimoto-book}.
The important problem here is that general principles of identifying the Langevin equation are not known.
Indeed, there are many examples of nonequilibrium Brownian motion that have not been described by the Langevin equations.

Thus, our aim in this paper
is to obtain a 
Langevin equation to reproduce the statistical properties of nonequilibrium Brownian motion,  particularly directional Brownian motion exhibiting a finite velocity on average.
We focus on higher cumulants of long-time displacement in the nonequilibrium Brownian motion and find that they work as a discriminator to identify the Langevin equation.
To demonstrate, we adopt a standard model, called the Rayleigh piston, which is used to derive the equilibrium Langevin equation from the microscopic dynamics of a large number of degrees of freedom \cite{vanKampen,Ford-Kac-Mazur}.
Its nonequilibrium version is presented in textbooks and has been studied during the past few decades \cite{Callen,Feynman-Leighton-Sands,Lieb,Gruber-Lesne,Gruber-Piasecki,Gruber-Frachebourg,Broeck,Munakata-Ogawa,Sinai,Meurs-Broeck-Garcia,Mansour-Garcia-Baras,Cencinietal,Fruleux-Kawai-Sekimoto,Sarrachino-Gnoli-Puglisi,Itami-Sasa1,Itami-Sasa2}.
We numerically identify a multiplicative Langevin equation (\ref{e:Langevin_neq_Ito}) that reproduces the cumulants of long-time displacement of this nonequilibrium Brownian motion at least up to the third order.
The multiplicative Langevin equation contains a linearly velocity-dependent friction coefficient, which can never appear in equilibrium.  

The paper is organized as follows. 
In section \ref{sec:setup}, we explain the setup of the Rayleigh piston model. 
In section \ref{sec:cum}, we define the cumulants of the long-time displacement and confirm that the third cumulant linearly increases with the observation time interval.
In section \ref{sec:effLan}, we identify the Langevin equation that can reproduce the cumulants up to at least the third order.
In section \ref{sec:rel}, we explain the relation between the original model and the identified Langevin equation, and then, we numerically estimate the scaling form of the cumulants in section \ref{sec:scaling}.
In section \ref{sec:stall}, we examine the dynamics of the stalled state and confirm that the property of the third cumulant is not affected.
The final section is devoted to a brief summary and some concluding remarks.

\section{Setup to generate nonequilibrium Brownian motion} \label{sec:setup}

\begin{figure}
\centering
\includegraphics[width=0.75\linewidth]{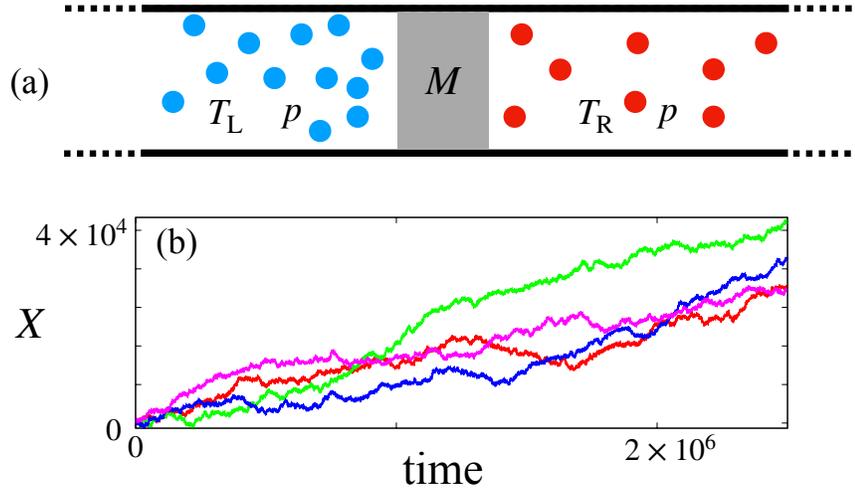}
\caption{(Color online) (a) Schematic illustration of  the Rayleigh piston driven by a temperature difference. 
(b) Typical time sequences of the Brownian motion generated from (\ref{e:master}). $M=10$, $m=0.01$, $p=10$, 
$\kB\TL=6.2$ and $\kB\TR=16.2$.
}
\label{fig:setup}
\end{figure}

We consider the experimental setup schematically shown in Fig.~\ref{fig:setup}(a).
A rigid piston of mass $M$ is freely movable in one direction inside an infinite cylinder of cross-sectional area $S$. 
Its position and velocity are denoted by $X$ and $V$, respectively.
The piston separates the cylinder into two regions filled with ideal gas particles of mass $m\ll M$.
The pressure $p$ is equal on both sides, whereas the temperature on the left side, $\TL$, is different from that on the right side, $\TR$.
Suppose, without loss of generality, that $\TL < \TR$.
For a later purpose, we define
\begin{eqnarray}
T &= \sqrt{\TL\TR},\\
\delta &= \frac{\TR-\TL}{T}.
\end{eqnarray}

Assuming that
the particles are in equilibrium before colliding with the piston and that they collide elastically and instantaneously with the piston only once,
we model the collisions between the piston and gas particles by random events  with a collision rate of
\begin{eqnarray}
\fl \lambda(v,V) = \frac{pS}{\kB\TL}(v-V)\theta(v-V) f^{\mathrm{L}}_{\mathrm{eq}}(v) 
+\frac{pS}{\kB\TR}(V-v)\theta(V-v) f^{\mathrm{R}}_{\mathrm{eq}}(v)
\end{eqnarray}
with
\begin{equation}
  f^\subLR_\eq(v)=\sqrt{\frac{m}{2\pi \kB T_\subLR}}\exp\left(\frac{-mv^2}{2\kB T_\subLR}\right),
\end{equation}
where $v$ is the velocity of a colliding particle, $\kB$ is the Boltzmann constant, $f^\subLR_\eq(v)$ is the Maxwell distribution, 
and $\theta(\cdot)$ is the Heaviside step function.
The coefficients \(p / k_BT_\subLR\) are equivalent to the number densities of the particles.
Using the laws of the conservation of energy and momentum, the transition probability per unit time from $V$ to $V'$, $W(V'\vert V)$, is given by 
\begin{eqnarray}
W(V'\vert V) = \lambda(v,V)\frac{\mathrm{d}v}{\mathrm{d}V'}
\end{eqnarray}
with 
\begin{equation}
  v=\frac{M+m}{2m}V'-\frac{M-m}{2m}V.
\end{equation}
Then, noting that $\dot{X}=V$, the time evolution of the probability density of $X$ and $V$ at time $t$, $P(X,V,t)$, is governed by the following master-Boltzmann equation:
\begin{eqnarray}
\frac{\partial P(X,V,t)}{\partial t} =& -V\frac{\partial P(X,V,t)}{\partial X}
 +\int\mathrm{d}V' W(V\vert V')P(X,V',t) 
 \nm
 &
-\int\mathrm{d}V' W(V'\vert V)P(X,V,t) .
\label{e:master}
\end{eqnarray}

Using the Gillespie algorithm~\cite{Gillespie}, 
we numerically obtain statistically correct trajectories of the master-Boltzmann equation (\ref{e:master}) without any approximations.

\section{Cumulants of the displacement as observables} \label{sec:cum}
As is done in typical experiments, we perform a time-lapse observation of an interval $t$ for a time series of $X$ numerically produced from (\ref{e:master}); see Fig.~\ref{fig:setup}(b).
We are interested in the long-time properties of the directional Brownian motion but not in the instantaneous velocity $V$, as it is not accessible by a time-lapse observation of $X$.
Rather, we concentrate on the displacement of $X$ in the interval $t$, 
\begin{eqnarray}
\Delta X_{t} \equiv X(t_0+t)-X(t_0).
\end{eqnarray}
The long-time fluctuations of the Brownian motion
are fully characterized by cumulants 
$\langle (\Delta X_{t})^{n}\rangle_{\mathrm{c}}$; more concretely, 
\begin{eqnarray}
\langle \Delta X_{t}\rangle_{\mathrm{c}} &=\langle \Delta X_{t}\rangle,\\
\langle (\Delta X_{t})^2\rangle_{\mathrm{c}} &=\langle (\Delta X_{t}-\langle \Delta X_{t}\rangle)^2\rangle,\\
\langle (\Delta X_{t})^3\rangle_{\mathrm{c}} &=\langle (\Delta X_{t}-\langle \Delta X_{t}\rangle)^3\rangle
\end{eqnarray}
 for $n=1,2,3$.
Here $t_0$-dependence can be ignored in the statistics for steady states. $\langle \cdot \rangle$ indicates a sample average and/or an average over $t_0$. 
The skewness is defined from these cumulants as $\langle (\Delta X_{t})^3\rangle_{\mathrm{c}}/(\langle (\Delta X_{t})^2\rangle_{\mathrm{c}})^{3/2}$.

\begin{figure}[bt]
\centering
\includegraphics[width=0.6\linewidth]{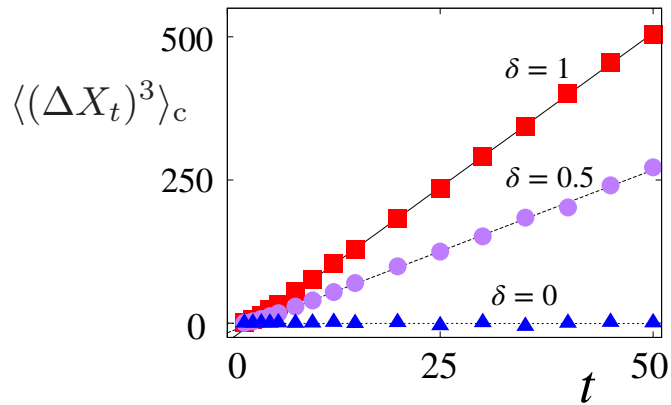}
\caption{Third cumulant of the displacement $\Delta X_t$ as a function of the observation interval $t$. The parameters of (\ref{e:master}) are $M=10$, $m=0.01$, $p=10$ and $\kB T=10$
for $\delta=0$, $0.5$, and $1.0$.
The three lines are linear fittings to the data for $\delta=0$, $0.5$, and $1.0$, respectively.
}
\label{fig:slope}
\end{figure}

These cumulants depend on the observation interval $t$.
For usual Brownian motion, the first or second cumulant is expected to increase linearly with the interval $t$ for sufficiently large $t$.
When the probability density of $\Delta X_t/t$ satisfies the large deviation property, $\langle (\Delta X_{t})^{n}\rangle_{\mathrm{c}}=O(t)$ as $t\rightarrow\infty$, except when $\langle (\Delta X_{t})^{n}\rangle_{\mathrm{c}}=0$ \cite{Bodineau-Derrida}.
We then focus on the growth rate of the $n$th cumulant 
\begin{eqnarray}
s_n\equiv \lim_{t\rightarrow \infty} \frac{\langle (\Delta X_{t})^{n}\rangle_{\mathrm{c}}}{t},
\label{e:slope}
\end{eqnarray}
where $s_1$ and $s_2/2$ 
are the mean velocity and the diffusion constant  for Brownian motion, respectively.
In real time sequences, we determine $s_n$ as the slope of cumulants around a finite value of $t$ such as $t\gg t_\mathrm{r}$ or typically $t\simeq 10 t_\mathrm{r}$, 
 where $t_\mathrm{r}$ is a characteristic relaxation time.

We demonstrate that the growth rate $s_3$ for the third cumulant can be finite in nonequilibrium Brownian motion.
Figure \ref{fig:slope} shows a linear increase in the third cumulant as a function of $t$
for the time sequences of $X$ produced by (\ref{e:master}).
We obtain  $s_3\neq 0$ out of equilibrium, $\TL\neq \TR$, and $s_3=0$ at equilibrium, $\TL=\TR$.
Thus, the third cumulant $\langle (\Delta X_{t})^{3}\rangle_{\mathrm{c}}$  
is never neglected for the nonequilibrium Brownian motion
even in the limit of sparse observation $t\rightarrow \infty$.

\section{Effective Langevin equation} \label{sec:effLan}

Hereafter, we focus on whether or not the nonequilibrium Brownian motion is described by a certain effective model.
The third cumulant works as a discriminator to determine the validity of various candidate models of the long-time behavior. For instance, a standard Langevin equation for a particle under a constant force $f$ with a constant friction coefficient,
$\gamma \dot X = f+\sqrt{2\gamma \kB T}\xi(t)$,
gives $s_3=0$ for any $f$; therefore, it is not appropriate as an effective model. The multiplicative Langevin equation with $\gamma(X)$ might become the next candidate, but this does not satisfy the translational invariance with respect to $X$. An underdamped Langevin equation for
$V= \dot X$, i.e., $ M\dot V = f-\gamma V+\sqrt{2\gamma \kB T}\xi(t)$,
with the mass $M$ of the object, also shows $s_3=0$.

\begin{figure}[bt]
\centering
\includegraphics[width=0.65\linewidth]{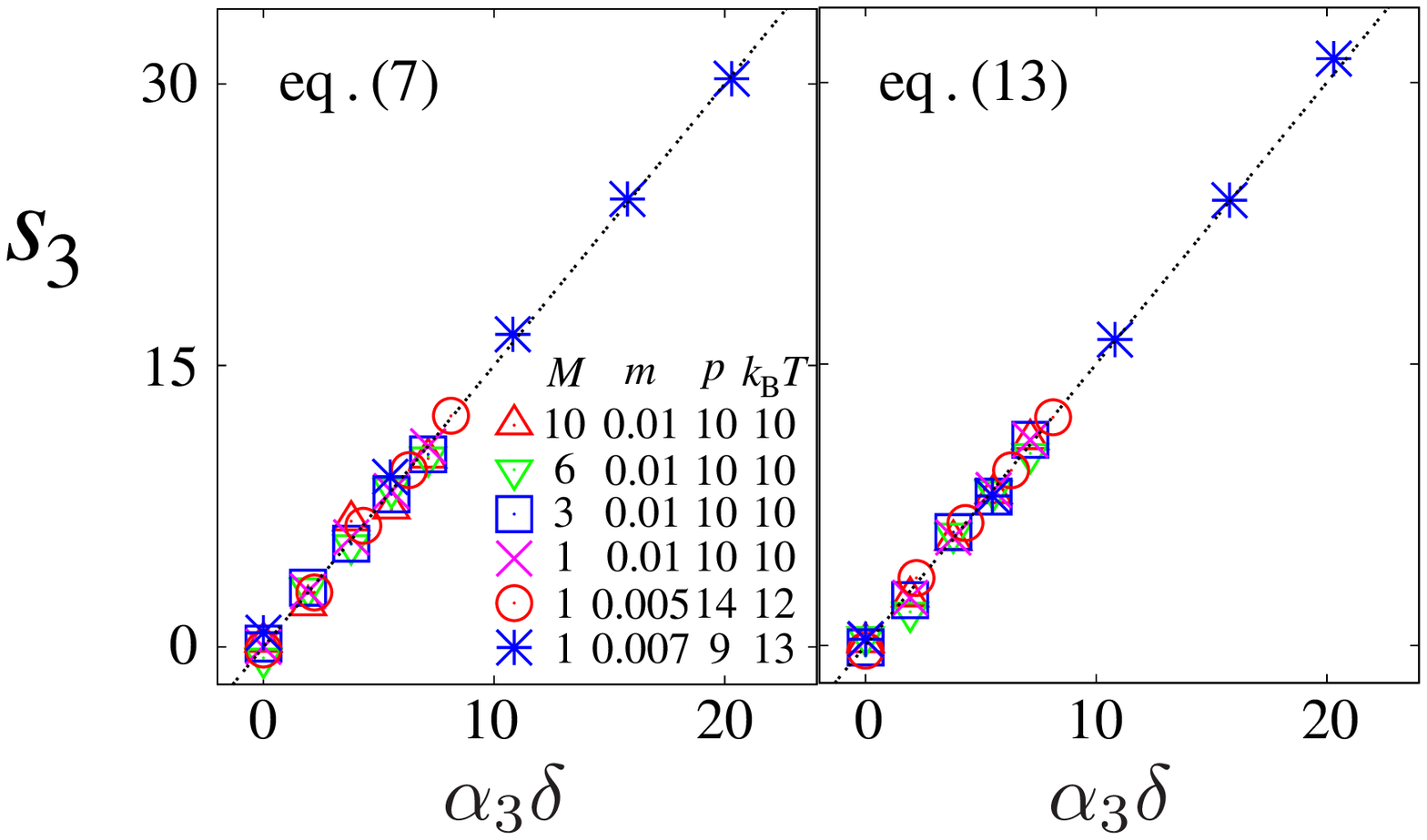}
\includegraphics[width=0.9\linewidth]{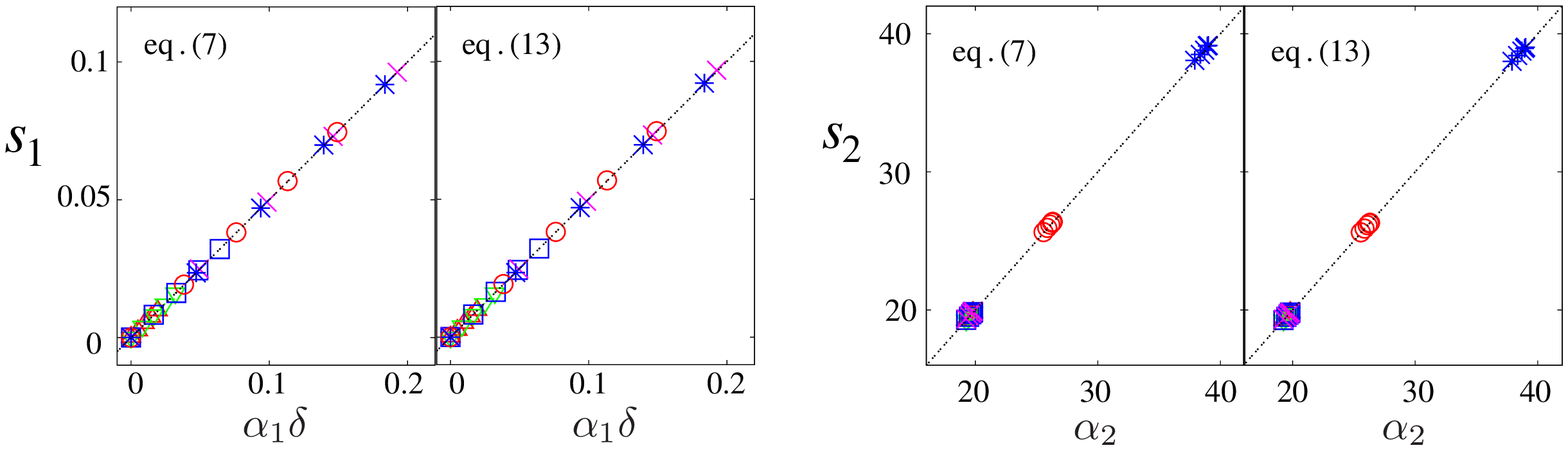}
\caption{ Coincidence of $s_1$, $s_2$ and $s_3$ for the time sequence generated from eq.~(\ref{e:master}) with those for the time sequence generated from 
 the Langevin equation (\ref{e:Langevin_neq_Ito}).
Each dotted line corresponds to eq.~(\ref{e:cum1}), (\ref{e:cum2}), or (\ref{e:cum3}),
i.e., $s_3=1.5 \alpha_3\delta$ with  $\alpha_3=\phi^3({\kB T}/{pS})^2\sqrt{{\kB T}/{m}}$,
$s_1=0.5\alpha_1\delta$ with $\alpha_1=\phi m/M\sqrt{{\kB T}/{m}}$, or
$s_2=\alpha_2$ with $\alpha_2=\phi {\kB T}/{pS} \sqrt{{\kB T}/{m}}$.
The parameters $(M, m, p, T)$ are given in the top-left section of the figure, whereas  $0\le \delta\le 1$.
}
\label{fig:collapse}
\end{figure}

The nonvanishing third cumulant, $s_3\neq 0$, does not imply the absence of the effective model.
For the directional Brownian motion generated from (\ref{e:master}), 
we identify the multiplicative Langevin equation 
 that reproduces the cumulants of the long-time displacement up to at least the third order. 
The equation is written as
\begin{eqnarray}
 M \dot{V} =& - \gamma(V) V + \sqrt{2\gamma(V) \kB T} \cdot \xi(t),
 \label{e:Langevin_neq_Ito}\\
 &\gamma(V)=\gamma_0(1-\gamma_1V)
\end{eqnarray}
where
the constants $\gamma_0$,  $\gamma_1$, and $T$ are determined by the system parameters. 
The symbol \(\cdot\) denotes the It\^o product, and
\(\xi\) is Gaussian white noise with zero mean and unit variance.
At the top of Figs.~\ref{fig:collapse}, the growth rate $s_3$ of the third cumulant obtained 
from the numerical integration of the effective Langevin equation (\ref{e:Langevin_neq_Ito}) using the Euler-Murayama method
is plotted parallel to $s_3$
obtained from the original Brownian motion of the Rayleigh piston (\ref{e:master}).
The points of the two figures coincide with each other within the statistical error.
The values of $s_1$ and $s_2$ are also consistent between (\ref{e:master}) and (\ref{e:Langevin_neq_Ito}), as shown at the bottom of Figs.~\ref{fig:collapse}.
Thus, we conclude that the multiplicative Langevin equation (\ref{e:Langevin_neq_Ito}) can be employed as an effective model for (\ref{e:master}) reproducing the long-time statistical properties.

In the long time sequence generated from (\ref{e:master}),
we determine the parameters as
\begin{eqnarray}
 \gamma_0=& 2\phi^{-1}pS \sqrt{\frac{m}{\kB T} },\label{e:gamma0}\\
 \gamma_1=&\delta pS\frac{m}{\gamma_0\kB T}\label{e:gamma1}
\end{eqnarray}
with
\begin{eqnarray}
 \phi= \sqrt{\frac{\pi}{8}}\frac{2\sqrt{T}}{\sqrt{\TL}+\sqrt{\TR}},\label{e:phi}
\end{eqnarray}
according to the argument in section \ref{sec:rel}.
The method to deduce the Langevin equation (\ref{e:Langevin_neq_Ito}) for general Brownian motion
from the time-lapsed observation is discussed in section \ref{sec:sum}.

\section{Relation between (\ref{e:master}) and (\ref{e:Langevin_neq_Ito})} \label{sec:rel}

Let us introduce a nondimensional small parameter 
\begin{eqnarray}
\ep \equiv \sqrt{\frac{m}{M}}\ll 1.
\end{eqnarray}
Using the Kramers--Moyal expansion and a perturbation expansion in powers of $\ep$ for (\ref{e:master}), we have~\cite{Gruber-Piasecki,AlkemadeETAL}
\begin{eqnarray}
\fl \frac{\partial P}{\partial t} =  -V\frac{\partial P}{\partial X} + \frac{\partial}{\partial V}\left( \frac{\gamma(V)V}{M}P\right) + \frac{\gamma_0 \kB T}{M^2}\frac{\partial^2 P}{\partial V^2}  + \frac{2\ep^2\delta  pS\kB T}{M^2} \frac{\partial^3 P}{\partial V^3}  + O(\ep^3),
\label{e:master_neq}
\end{eqnarray}
which is explained in the Appendix.
 Here, $\gamma_0$ and $\gamma_1$ are given by (\ref{e:gamma0}) and (\ref{e:gamma1}), respectively.
$T=\sqrt{\TL\TR}$ corresponds to the kinetic temperature of the piston,
i.e., $M\langle V^2\rangle_{\rm c}=\kB \sqrt{\TL\TR}+O(\ep^2)$ 
\cite{Gruber-Lesne,Gruber-Piasecki}.

Ignoring the contribution of $O(\ep^2)$ by noting that both $\gamma_0$ and $\gamma_1$ are of $O(\ep)$, (\ref{e:master_neq}) corresponds to a Langevin equation 
\begin{eqnarray}
M\dot{V} = -\gamma_{0} V + \sqrt{2 \gamma_0 \kB T} \xi(t),
\label{e:lowest}
\end{eqnarray}
which cannot reproduce the directional motion and $s_1=0$.
Thus, we cannot ignore the $O(\ep^2)$ terms. 
Including the contribution of $O(\ep^2)$, 
(\ref{e:master_neq}) becomes similar to the Fokker--Planck equation but still contains the third derivative term,
which implies that the effective dynamics for the nonequilibrium Brownian motion of (\ref{e:master}) is not
straightforwardly concluded from the Kramers--Moyal expansion.

Suppose that (\ref{e:master_neq}) is transformed into
\begin{eqnarray}
\frac{\partial P}{\partial t} =  -V\frac{\partial P}{\partial X} + \frac{\partial}{\partial V}\left( \frac{\gamma(V)V}{M}P\right) + \frac{\kB T}{M}\frac{\partial^2 }{\partial V^2} \left( \frac{\gamma(V)}{M}  P\right)
+ O(\ep^3)
\label{e:renormal}
\end{eqnarray}
by coarse-graining in time.
The expression (\ref{e:renormal}) is consistent with the multiplicative Langevin (\ref{e:Langevin_neq_Ito}).
Although we do not know the procedure to derive (\ref{e:renormal}), we know it is appropriate from the numerical examination shown in Figs.~\ref{fig:collapse}.
The Langevin-like equation derived in \cite{Plyukhin-Schofield} is not compared with (\ref{e:renormal}) because the statistical properties of the noise are not obvious.

\section{ Scaling form for the cumulants} \label{sec:scaling}

Hereafter, we explain a scaling form for the cumulants that plays a central role in identifying the effective Langevin equation (\ref{e:Langevin_neq_Ito}) for any parameter value in $\ep \ll 1$.
We introduce the rescaled dimensionless variables 
\begin{eqnarray}
\tau &\equiv \frac{t}{t_\mathrm{r}},\label{e:tau}\\
\mathcal{V} &\equiv \frac{V}{V_{\mathrm{th}}},\label{e:calV}\\
\mathcal{X} &\equiv \frac{X}{V_{\mathrm{th}}t_\mathrm{r}}\label{e:calX}
\end{eqnarray}
 to summarize the numerical experiments.
The variables satisfy \(\mathcal{V} = \mathrm{d}\mathcal{X}/\mathrm{d}\tau\).
Here,
$t_\mathrm{r}=M/\gamma_0$ from (\ref{e:lowest}), 
\(V_\mathrm{th} \equiv\sqrt{\kB T / M}\)  is the effective thermal velocity 
and \(V_\mathrm{th} t_\mathrm{r}\) is the characteristic length scale.
The probability density of $\mathcal{X}$ and $\mathcal{V}$ at $\tau$, $\mathcal{P}(\mathcal{X},\mathcal{V},\tau)$, is given by 
\begin{eqnarray}
\mathcal{P}(\mathcal{X},\mathcal{V},\tau) = P(X,V,t)\frac{\mathrm{d}X}{\mathrm{d}\mathcal{X}}\frac{\mathrm{d}V}{\mathrm{d}\mathcal{V}}
\end{eqnarray}
Then, using these new variables, (\ref{e:master_neq}) is rewritten as
\begin{eqnarray}
\frac{\partial \mathcal{P}}{\partial \tau} = -\mathcal{V}\frac{\partial\mathcal{P}}{\partial\mathcal{X}} + \frac{\partial}{\partial\mathcal{V}}\Big((1-\ep g_1)\mathcal{V}\mathcal{P}\Big) + \frac{\partial^{2}\mathcal{P}}{\partial\mathcal{V}^{2}}+ 2\ep g_1
\frac{\partial^{3}\mathcal{P}}{\partial\mathcal{V}^{3}}  +O(\ep^{2})
\label{e:master_nondim}
\end{eqnarray}
with
\begin{eqnarray}
g_1 = \sqrt{\frac{\pi}{8}}
\left[\left(\frac{\TR}{\TL} \right)^{\frac{1}{4}}-\left(\frac{\TL}{\TR}\right)^{\frac{1}{4}}\right],
\end{eqnarray}
as derived  in the Appendix.
Here,
$g_1\simeq\sqrt{{\pi}/{32}}\delta$ for $0\le \delta \le1$.

We first simulate (\ref{e:master}) for $(M,m,T,p,S)=(10,0.01,10,10,1)$ and $0\le \delta\le 1$, and determine the scaled version of growth rates
 \(\langle (\Delta \mathcal{X}_\tau)^n\rangle_{\rm c} / \tau\)  as a function of $g_1$ for sufficiently long $\tau$.
Numerical fitting by the least-squares method gives estimates for $n=1, 2, 3$  as
\begin{eqnarray}
\frac{\langle \Delta \mathcal{X}_{\tau}\rangle_{\mathrm{c}}}{\tau} &= (1.002 \pm 0.009)\ep g_1,\label{e:calS1}\\
\frac{\langle (\Delta \mathcal{X}_{\tau})^{2}\rangle_{\mathrm{c}}}{\tau} &= 2.0006 \pm 0.0005,\label{e:calS2}\\
\frac{\langle (\Delta \mathcal{X}_{\tau})^{3}\rangle_{\mathrm{c}}}{\tau} &= (12.02 \pm 0.06)\ep g_1\label{e:calS3}
\end{eqnarray}
with asymptotic standard errors.
Rewriting the above fits in dimensional forms, we have estimates of
\begin{eqnarray}
s_1 &=\frac{\langle \Delta X_t\rangle}{t} =
\frac{\phi}{2}
 \sqrt{\frac{\kB T}{m}}\frac{m}{M}\delta,\label{e:cum1}\\
s_2 &=\frac{\langle (\Delta X_t)^2\rangle_{\mathrm{c}}}{t} =
\phi\sqrt{\frac{\kB T}{m}}\frac{\kB T}{pS},\label{e:cum2}\\
s_3 &=\frac{\langle (\Delta X_t)^3\rangle_{\mathrm{c}}}{t} =
\frac{3\phi^3}{2}
\sqrt{\frac{\kB T}{m}}\left(\frac{\kB T}{pS}\right)^2\delta,
\label{e:cum3}
\end{eqnarray}
where $t\gg t_\mathrm{r}$.
Next, we change the parameters $M$, $m$, $p$, $\TL$ and $\TR$ and simulate  (\ref{e:master}) while keeping $S=1$ without loss of generality.
We plot the numerical results of the growth rates $s_1$, $s_2$ and $s_3$ 
in Figs. \ref{fig:collapse}.
It is remarkable that all the data are collapsed in each line corresponding to (\ref{e:cum1}), (\ref{e:cum2}) or (\ref{e:cum3}) for $0.032\le \ep \le 0.141$ and $0\le \delta \le 1$.
Thus, 
we can determine the value of the growth rates $s_n$ for any parameter value.
Note that (\ref{e:calS1}), (\ref{e:calS2}), and (\ref{e:calS3})  depend on only $\ep$ and $g_1$, consistent with
the parameters included in  (\ref{e:master_nondim}), without $O(\ep^2)$ errors.
This confirms that the contribution of $O(\ep^2)$ in (\ref{e:master_nondim})  does not affect the Brownian motion for $\ep\ll 1$.

From (\ref{e:cum1}), (\ref{e:cum2}), and (\ref{e:cum3}),  we obtain
the mean displacement as of $O(t \delta)$, the variance of the displacement as $O(t)$ and
the skewness of the displacement as $O(t^{-1/2} \delta)$.
Thus, the skewness vanishes in the long-time limit $t\rightarrow \infty$,
and then the displacement of the Brownian motion may be considered to obey a Gaussian distribution.
However, we cannot draw the Langevin equation (\ref{e:Langevin_neq_Ito})
starting from the Gaussian distribution.
This indicates that the third cumulant is superior to skewness in examining the long-time dynamics.

\section{Nonequilibrium Brownian motion without directionality}
\label{sec:stall}

Here, we compare the Langevin equation (\ref{e:Langevin_neq_Ito}) with the general form of the Langevin equation in equilibrium systems.
In an equilibrium system,
when the stationary distribution of $V$ is 
\begin{eqnarray}
 P(V)=\sqrt{\frac{M}{2\pi \kB T}}\exp\left( -\frac{MV^{2}}{2\kB T}\right),
\end{eqnarray}
a multiplicative Langevin equation for $V$ is generally given by
\begin{equation}
 M\dot{V}=-\gamma_{\mathrm{eq}}(V)V+\sqrt{2\gamma_{\mathrm{eq}}(V)\kB T}\odot \xi(t),
\label{e:L_eq}
\end{equation}
 where $\odot$ denotes the anti-It\^o product.
The nonlinear friction coefficient $\gamma_{\mathrm{eq}}(V)$ is limited to the time-symmetric form, i.e., $\gamma_{\mathrm{eq}}(V) = \gamma_{\mathrm{eq}}(-V)$,
when the detailed balance condition or time-reversibility is imposed ~\cite{Graham-Haken,Graham-Haken2}.
To compare the Langevin equation (\ref{e:Langevin_neq_Ito}) with the standard form (\ref{e:L_eq}), we change the product of (\ref{e:Langevin_neq_Ito}) from It\^o to anti-It\^o.
We obtain
\begin{eqnarray}
 M\dot{V}= \ep^{2}\delta pS-\gamma(V) V+\sqrt{2\gamma(V) \kB T}\odot \xi(t),
\label{e:Langevin_neq_aIto}
\end{eqnarray}
which is essentially different from the standard form (\ref{e:L_eq}) in two aspects: the constant force term $\ep^{2}\delta pS$ and the time-irreversibility of the friction coefficient, 
$\gamma(V)\neq\gamma(-V)$.

\begin{figure}
\centering
\includegraphics[width=0.75\linewidth]{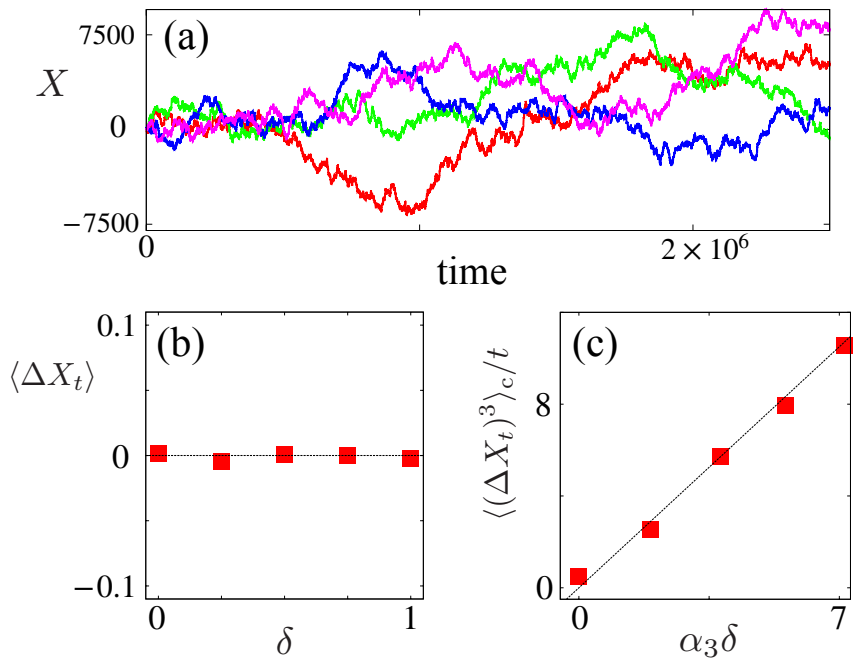}
\caption{(Color online) The dynamics of the piston under the stall force
$-\ep^2\delta pS$. 
$M=10$, $m=0.01$, $p=10$, $\kB T=10$ and $0\le \delta\le 1$.
(a) Time sequences of the piston when $\delta=1.0$.
(b) $\langle \Delta X_t\rangle$ for the observation interval $t=100$.
(c) $\langle (\Delta X_t)^3\rangle_{\rm c}/t$ as a function of $\alpha_3 \delta$. The line is $\langle (\Delta X_t)^3\rangle_{\rm c}/t=1.5 \alpha_3\delta$.}
\label{fig:pull}
\end{figure}

We notice that the effective Langevin equation  
looks similar to (\ref{e:L_eq}) when a drift term $-\ep^{2}\delta pS$ is added.
Correspondingly, the piston with a constant force $-\ep^{2}\delta pS$ exhibits rather typical Brownian motion without the mean displacement, as shown in Fig.~\ref{fig:pull}(a).
The statistical average in Fig.~\ref{fig:pull}(b) indicates 
$\langle V\rangle=0$, as expected.
In this sense, $\ep^{2}\delta pS$ in (\ref{e:Langevin_neq_aIto}) may be regarded as a force generated by the temperature difference.
Even in the stalled state with $\langle V\rangle=0$, 
the dynamics of the piston are not equivalent to  Brownian motion satisfying detailed balance, 
as
the third cumulant $\langle(\Delta X_t)^3\rangle_{\rm c}$ 
is scaled in the same manner as in (\ref{e:cum3}); see Fig.~\ref{fig:pull}(c).
This non-Gaussian nature is due to the time-asymmetric friction coefficient $\gamma(V)\neq\gamma(-V)$.

\section{Concluding remarks} \label{sec:sum}

We have proposed a new principle to identify the Langevin equation describing nonequilibrium Brownian motion by focusing on the statistical properties of long-time displacement.
To demonstrate our new principle, we have studied the nonequilibrium version of the Rayleigh piston model (\ref{e:master}) and have identified the Langevin equation (\ref{e:Langevin_neq_Ito}) that can reproduce the cumulants up to at least the third order.
The identified Langevin equation involves the velocity-dependent friction coefficient 
 $\gamma(V)\neq\gamma(-V)$, which breaks the time-reversibility.
This leads to a breaking of the detailed balance and the non-Gaussian nature of the Brownian motion; moreover, it leads to the divergence of the entropy production 
defined by a log ratio of the weight for a trajectory and its time reverse.
Thus, (\ref{e:Langevin_neq_Ito}) can reproduce the long-time fluctuations,
but we need another viewpoint to investigate the thermodynamic properties of (\ref{e:master}).

Our results show that the nonequilibrium Rayleigh piston provides a good model for theoretical statistical mechanics because of the following two reasons: First, our numerical study ensures that the nonequilibrium Rayleigh piston is described by the Langevin equation at a coarse-grained scale.
Second, the expression of the Langevin equation (\ref{e:Langevin_neq_Ito}) is not straightforwardly deduced from the equilibrium expression. 
The newly found Langevin equation contains a friction coefficient that is completely different from the friction coefficient in equilibrium systems and that has not been considered in previous studies.  
Our expression stimulates theoretical studies to derive the Langevin equation and moreover opens the discussion to determine a standard form for nonequilibrium Brownian motion. 
At present, even the growth rate of the higher cumulants is difficult to determine.
In the Nos\'e-Hoover equation of motion with non-constant friction coefficients, higher cumulants of the velocity fluctuations have been calculated \cite{Hoover-Holian}. 
By examining this example, we may find a systematic method for calculating higher cumulants under general nonequilibrium setups.

Before ending this paper, we make some remarks.
Our method may be applied to a general time sequence and may then stimulate various fields such as colloidal particles, biomotors in cells, agents in active matter, stars within galaxies, interfaces, market prices and so on. 
We emphasize that the experimental trial does not have an associated cost: 
it does not require expensive equipment of high time resolution and memorizes only the long time sequences.
We wonder if a mechanism, i.e.,
 the time asymmetry of the friction coefficient generating an apparent force,
occurs in a large variety of directional Brownian motion in nature.
We hope that the present experimental analysis is relevant to the time series of various cases of directional Brownian motion.
The most important point is to determine the dependence of the third cumulant for the displacement as a function of the observation time interval.
Once we obtain this dependence, the friction coefficients may be deduced heuristically.
In what follows, we try to propose a scheme to deduce the friction coefficients and Langevin equations.
Suppose that $M$ and $T$ are known from other experiments.
First, we assume the form of the Langevin equation as
\begin{eqnarray}
 M \dot{V} = f- \gamma_0(1-\gamma_1 V) V + \sqrt{2\gamma_0(1-\gamma_1 V) \kB T} \cdot \xi(t),
 \label{e:general_form}
\end{eqnarray}
that is, we assume that a fluctuation-dissipation-like theorem holds even in nonequilibrium. 
In (\ref{e:general_form}), we adopted the It\^o product without loss of generality, as the change in the product generates only a constant force absorbed into $f$.
Next, we calculate  the cumulants of the displacement for a time interval $t$
by solving (\ref{e:general_form}) analytically or numerically for various $\gamma_1$ and $f$ and
denote the cumulants as $\langle (\Delta X_t)^n\rangle_{\rm c}^{\gamma_1,f}$, specifying $\gamma_1$ and $f$ in (\ref{e:general_form}).
Finally, we obtain simultaneous equations for $\gamma_1$ and $f$ such that
\begin{eqnarray}
\lim_{t\rightarrow\infty}\frac{\langle (\Delta X_t)^n\rangle_{\rm c}^{\gamma_1,f}}{t}=s_n, 
\quad (n=1,2,3)
\label{e:deduce}
\end{eqnarray}
where $s_n$ denotes the growth rates of the $n$th cumulant  in (\ref{e:slope}) determined from the experimental time sequence.
By solving these equations, we arrive at estimates for $\gamma_1$ and $f$.
If the simultaneous equations (\ref{e:deduce}) do not have a solution, the time sequence may not be explained by (\ref{e:Langevin_neq_Ito}) or (\ref{e:general_form}) but belongs to another statistical category.
We believe that our findings will lead to a fundamental change in the standard form of nonequilibrium Brownian motion.

\ack
The authors would like to thank Shin-ichi Sasa for perpetual stimulating discussions and useful comments.
The present study was supported by JSPS KAKENHI Grant Numbers JP15K05196, JP17K14355, JP17K14373, JP17H01148 , JP19K03647 and JP19H01864.

\appendix
\setcounter{section}{1}
\section*{Appendix : Derivations of (\ref{e:master_neq}) and (\ref{e:master_nondim})}

Applying the Kramers--Moyal expansion to  the master-Boltzmann equation (\ref{e:master}), we obtain
\begin{eqnarray}
\fl \frac{\partial P(X,V,t)}{\partial t} = & -V\frac{\partial P}{\partial X}
 \nm
 & +\sum^{\infty}_{k=1} \frac{(-1)^k}{k!} \frac{\partial ^ k}{\partial V^k}   \left ( \frac{2m}{M+m} \right )^k   \int_{-\infty}^{\infty} dv \lambda(v,V)(v-V) ^k P(X,V,t).
\label{masterKur}
\end{eqnarray}
By rescaling the variables according to (\ref{e:tau}), (\ref{e:calV}) and (\ref{e:calX})
as $t\rightarrow \tau$, $X\rightarrow \mathcal X$, and $V\rightarrow \mathcal V$,
(\ref{masterKur})  is rewritten in dimensionless form as
\begin{eqnarray}
\fl \frac{\partial {\mathcal P}(\mathcal X, \mathcal V,\tau)}{\partial \tau} 
=  -\mathcal V\frac{\partial \mathcal P}{\partial \mathcal X}
\nm
 +\frac{1}{4} \frac{1}{A+A^{-1}}  \sum^{\infty}_{k=1} \frac{(-1)^k}{k!}\frac{\partial^k}{\partial \mathcal V ^k} \left ( \frac{2 \ep^2}{1+\ep^2}  \right)^k 
\int_{-\infty}^{\infty} d \tilde v~ \tilde \lambda(\tilde v,\mathcal V)  (\tilde v - \mathcal V) ^k {\mathcal P}(\mathcal X,\mathcal V,\tau) ,
\nm
\label{master_nondim_full}
\end{eqnarray}
where 
\begin{eqnarray}
\tilde \lambda( \tilde v , \mathcal V) 
&:=   
A^{-3}  (\tilde v- \mathcal V )\theta(\tilde v-\mathcal V)  e^{\frac{- \ep^2 }{2} \frac{\tilde v^2}{A^2}} 
+ A^3   (\mathcal V- \tilde v)\theta( \mathcal V- \tilde v)   e^{\frac{- \ep^2 }{2}A^2\tilde v^2}
 \end{eqnarray}
and 
\begin{eqnarray}
A\equiv\left(\frac{\TL}{\TR}\right)^{\frac{1}{4}}.
\end{eqnarray}
We further transform the integral of (\ref{master_nondim_full}) as
\begin{eqnarray}
\fl \int_{-\infty}^{\infty} d \tilde v ~ \tilde \lambda(\tilde v,\mathcal V) (\tilde v -\mathcal V)^k  
 =  &\int^{\infty}_{-\infty} d u~ \left\{
A^{k-1} (u - A^{-1}{\mathcal V} )^{k+1}\theta(u - A^{-1}{\mathcal V} )
 \right.
 \nm
& \left.- A^{1-k}  (u - A\mathcal V )^{k+1} \theta(A\mathcal V - u) \right \} e^{\frac{- \ep^2 }{2}  u^2}.
\end{eqnarray}
Applying the perturbative expansion in $\ep$ and performing the integration, we have
\begin{eqnarray}
 \left ( \frac{2 \ep^2}{1+\ep^2} \right )^{k}\int d \tilde v \tilde \lambda(\tilde v,\mathcal V) (\tilde v - \mathcal V) ^{k}
 =O(\ep^{k-2})
\end{eqnarray}
for $k\ge 2$.
The explicit forms for $1\le k\le 5$ are calculated as
\begin{eqnarray}
\fl \left ( \frac{2 \ep^2}{1+\ep^2} \right ) \int d \tilde v~ \tilde \lambda(\tilde v,\mathcal V)   (\tilde v - \mathcal V) = -4(A+A^{-1}) \mathcal V - \sqrt{2\pi}(A^{2}-A^{-2})\mathcal V^2 \ep +O(\ep^2), \\
\fl \left ( \frac{2 \ep^2}{1+\ep^2} \right )^2\int d \tilde v~ \tilde \lambda(\tilde v,\mathcal V) (\tilde v - \mathcal V) ^2 = 8(A+A^{-1}) + O(\ep^2), \\
\fl \left ( \frac{2 \ep^2}{1+\ep^2} \right )^3\int d \tilde v~ \tilde \lambda(\tilde v,\mathcal V) (\tilde v - \mathcal V) ^3 = 12 \sqrt{2 \pi}(A^2 -A^{-2}  )\ep +O(\ep^2).
\end{eqnarray}
Substituting these evaluations into (\ref{master_nondim_full}), we obtain 
(\ref{e:master_nondim}), that is
\begin{eqnarray}
\fl \frac{\partial \mathcal P(\mathcal X,\mathcal V,\tau)}{\partial \tau} =& -\mathcal V\frac{\partial \mathcal P}{\partial \mathcal X} +\frac{\partial}{\partial \mathcal V }\left [ 1 -\sqrt{\frac{\pi}{8}} \left \{ \left( \frac{\TR}{\TL} \right)^{\frac{1}{4}} - \left( \frac{\TL}{\TR} \right )^{\frac{1}{4}} \right \} \ep \mathcal V \right ] \mathcal V \mathcal P(\mathcal X,\mathcal V,\tau) \nm 
 &+  \frac{\partial^2}{\partial \mathcal V^2 }\mathcal P(\mathcal X,\mathcal V,\tau) 
 \nm
 &+2 \sqrt{\frac{\pi}{8}} \left \{ \left( \frac{\TR}{\TL} \right)^{\frac{1}{4}} - \left( \frac{\TL}{\TR} \right )^{\frac{1}{4}} \right \} \ep  \frac{\partial^3}{\partial \mathcal V^3 }P(\mathcal X,\mathcal V,\tau) +O(\ep^2).
\label{Master_nondim_ex}
\end{eqnarray}
The transformation of (\ref{Master_nondim_ex}) into the dimensional form
leads to (\ref{e:master_neq}).

\end{document}